# Nanoscale Horizons

## COMMUNICATION

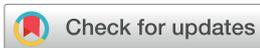 Check for updates

## Coupling of plasmonic hot spots with shurikens for superchiral SERS-based enantiomer recognition†

Olga Guselnikova, [ORCID] *[ac] Roman Elashnikov, [ORCID] [a] Vaclav Svorcik,[a] Martin Kartau,[b] Cameron Gilroy,[b] Nikolaj Gadegaard, [ORCID] [d] Malcolm Kadodwala, [ORCID] [b] Affar S. Karimullah [ORCID] *[b] and Oleksiy Lyutakov [ORCID] *[a]

Detection of enantiomers is a challenging problem in drug development as well as environmental and food quality monitoring where traditional optical detection methods suffer from low signals and sensitivity. Application of surface enhanced Raman scattering (SERS) for enantiomeric discrimination is a powerful approach for the analysis of optically active small organic or large biomolecules. In this work, we proposed the coupling of disposable chiral plasmonic shurikens supporting the chiral near-field distribution with SERS active silver nanoclusters for enantio-selective sensing. As a result of the plasmonic coupling, significant difference in SERS response of optically active analytes is observed. The observations are studied by numerical simulations and it is hypothesized that the silver particles are being excited by superchiral fields generated at the surface inducing additional polarizations in the probe molecules. The plasmon coupling phenomena was found to be extremely sensitive to slight variations in shuriken geometry, silver nanostructured layer parameters, and SERS excitation wavelength(s). Designed structures were able to discriminate cysteine enantiomers at concentrations in the nanomolar range and probe biomolecular chirality, using a common Raman spectrometer within several minutes. The combination of disposable plasmonic substrates with specific near-field polarization can make the SERS enantiomer discrimination a commonly available technique using standard Raman spectrometers.

### New concepts

Detection of enantiomers is a challenging problem for the pharmaceutical industry, clinical testing, monitoring of the environment, and food quality testing. Through the optical methods, the most common surface-enhanced Raman optical activity is often limited by complicated interpretation because a locally enhanced electric field has linear polarization and cannot selectively interact with circularly polarized light or different enantiomers. We show a new type of sensor based on the use of chiral plasmonic nanostructures – shurikens for surface enhanced Raman spectroscopy (SERS), which can generate chiral plasmon using linearly polarized light. Gold-coated shurikens were coupled with an array of Ag clusters, which are exposed to the superchiral field close to shurikens, followed by the expected excitation of circular polarization in plasmonic silver hot spots. We optimized the geometry of shurikens, Ag layer thickness/roughness and excitation wavelength to enantioselectively detect L/D-cysteine, L/D-methionine and (R)/(S)-thalidomide and DNA. This concept differs from other SERS-based chiral sensors, which utilize enantioselective adsorption of chiral analyte rather than the interaction between near-chiral field and molecule. The novel observation reported a broad impact, it designates new plasmonic materials in chiral sensing.

## Introduction

Chiral molecules are widespread in nature including in our bodies and various pharmaceutical drugs that we use.

[a] Department of Solid State Engineering, University of Chemistry and Technology, 16628 Prague, Czech Republic. E-mail: lyutakoo@vscht.cz
[b] School of Chemistry, Joseph Black Building, University of Glasgow, Glasgow, G12 8QQ, UK. E-mail: Affar.Karimullah@glasgow.ac.uk
[c] Research School of Chemistry and Applied Biomedical Sciences, Tomsk Polytechnic University, Tomsk, 634050, Russian Federation. E-mail: guselnikovaoa@tpu.ru
[d] James Watt School of Engineering, University of Glasgow, Rankine Building, Glasgow, G12 8LT, UK

Depending on their chirality, molecules can show significantly different pharmaceutical activity. Hence, the development of simple and universal methods of enantioselective discrimination is highly sought after.[1,2] Commonly used circular dichroism (CD) and optical rotation dispersion (ORD) spectroscopy represents universal techniques that can be applied to a range of molecules, starting from small organic enantiomers discrimination up to protein structure identification.[3–6] However, these methods do not provide detailed molecular-level information about the analyte which are susceptible to salt and solvent interferences which significantly restricts their application.[7,8] As an alternative Raman optical activity (ROA) is also widely used, providing information about the chemical structure of the analyte as well as its chirality.[9,10] However, the drawback of ROA spectroscopy is the weakness of the signal leading to unacceptably long measurement times.[11,12] To compensate this drawback, the utilization of a plasmon-based surface-enhanced ROA (SEROA) approach was proposed.[13–15] Indeed, sub-diffraction light focusing, ensured by surface plasmon can significantly enhance the Raman signal and accelerate the spectra collection.[16–19]

It was recently demonstrated that the utilization of plasmonics in enantiomeric discrimination is not restricted solely to SEROA.[20–22] As an example, firstly observed by Govorov *et al.*, gigantic plasmon-assisted enhancement of CD signals can also be obtained.[23] Discrimination of large biomolecule conformations were also performed using ROA with the utilization of plasmon-active metamaterials.[24–26] Additionally, plasmon-active structures were functionalized with secondary probe molecules to mediate stereospecific analyte–probe interactions or ensure selective entrapping of enantiomers near plasmon active surface for subsequent SERS measurements.[27–30]

On other hand, the significant progress in the field of nanomaterials construction and preparation led to gradual appearance of structures, able to support the excitation of intrinsically chiral plasmons.[9,31–34] Recent research indicate that such structures are promising candidate for SERS based enantiomeric discrimination.[32,33,35,36] The unique interaction of chiral plasmons and analyte molecules can selectively enhance the SERS spectrum of one enantiomer, making the SERS enantio-selective discrimination possible.[33,35,37,38] The precise mechanism of this approach remains unclear and chiral plasmon assisted label-free SERS is actually just at its infancy. However, such label free estimation of molecular structure and chirality can certainly be considered to have a promising future.

In this work, we propose the coupling of chiral shuriken-shaped plasmonic structures with plasmonic hot spots, originating from deposited silver clusters for SERS-based enantiomer recognition.[26] SERS measurements indicate the chiral-dependent intensity of the scattered light from the enantiomer probes, where the SERS intensity was determined by the handedness of the shuriken and the probe. This work demonstrates the use of replicable and customizable chiral nanopatterned surfaces as a simplified method of rapid enantiomeric discrimination with molecular level spectroscopic information.

# Results and discussion

## Design of optically and SERS-active structures

The nano-structures for SERS-based enantiomer discrimination are described in Fig. 1(a). As the chiral support, we used the previously developed nanostructures termed "shurikens" (SEM images with typical morphology in Fig. 1(b) and Fig. S1–S3, ESI†) having a six-fold rotational symmetry, either left-handed (LH) or right-handed (RH), placed in a square lattice to produce an array.[19,24–26,34,36] The samples were prepared using injection molding and coated with a Au film to produce the final metafilms. The shurikens generate fields with large chiral hotspots and overall net chirality at the metal–dielectric boundary (Fig. S4 and S5, ESI†). The coupling of magnetic and electric modes excited across the illuminated shuriken structures generate superchiral near fields that display greater chiral asymmetry than the circularly polarized plane waves (detailed description is given in previous works).[39–41]

The metafilms were sputter coated with Ag which causes a very subtle broadening of the resonances indicated in the reflectivity spectra, Fig. S4a (ESI†). Furthermore, in the samples used here, there are 6 shuriken designs with subtle difference, and the arrays for each design are labelled S1–S6 with '/Ag' indicating they are coated with Ag particles. Each design varies in two parameters for their initial master shim (used for injection moulding) design. First is the arm width and second is the diameter of a circle in the middle of shuriken (see Fig. S1–S4, ESI† and related discussion).

The differential interaction of LH and RH superchiral fields with surrounding enantiomers (for example, deposited as an additional layer) allows for their enantioselective sensing, as was demonstrated previously.[19,40,42,43] The shuriken structures used in this work were also previously employed for protein

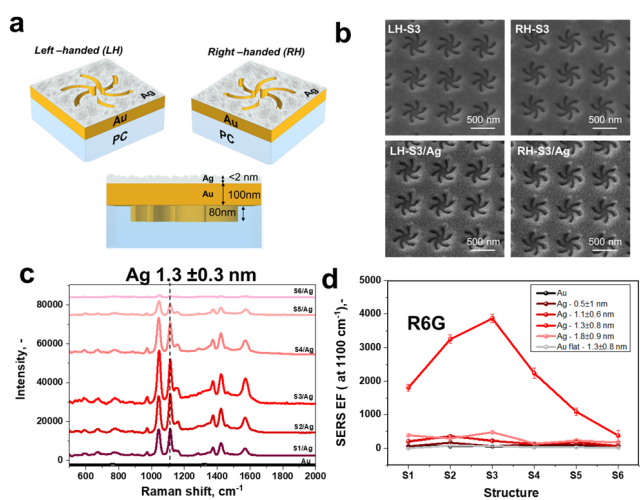

Fig. 1 (a) Schematic description of plasmon-active SERS substrates with coupled chiral active shuriken and cluster-like Ag layer; (b) SEM images of LH- and RH-handled shuriken structures before and after Ag deposition; (c) SERS spectra of R6G, measured at 785 nm on RH-S*x*/Ag structures as a function of shurikens geometry; (d) dependence of SERS enhancement factors on shurikens geometry and silver effective thickness.

secondary structure sensing through the measurement of the shifts of the resonance wavelength position or light phase retardation.[34,44] When directly attempting SERS measurements with just the pristine Au shuriken metafilms, no significant SERS enhancement was observed (Fig. 1(c)). Hence, in order to introduce additional plasmonic hot spots, we sputtered an additional silver layer with a cluster-like nature (Fig. S6–S10, ESI† and related discussions). The SERS enhancement can then be expected due to the rough nature of the deposited film.[45,46] The deposited silver clusters can be approximated as an array of plasmonic dipoles, which commonly ensure the high degree of SERS enhancement. To distinguish the impact of Ag for the material selection of the nanoparticles we compared it to Au nanoparticle only coatings on the Au shurikens to provide a highly roughened Au film. Fig. S11 (ESI†) shows that the S3/Ag has a major increase in SERS in comparison to Au only and the S3 with Au nanoparticles. Furthermore, Ag nanoparticles on their own also show much lower SERS enhancement (Fig. S12, ESI† and as Fig. 1 shows, the design of shuriken and the resonance feature impacts the overall SERS enhancement significantly.

The final structure (Fig. 1(a)) represents the Au shuriken coupled with Ag nanoclusters, hence, producing a surface of randomly oriented plasmonic dipoles exposed to the super-chiral optical near fields generated by the plasmonic shurikens. Furthermore, created samples are referred to as LH- or RH-S$x$/Ag, where LH- or RH-describes the shuriken left or right-handed rotation symmetry, S$x$ indicates the shuriken geometry (Fig. S1 and S2, ESI†) and Ag highlights the presence of added silver nanoclusters.

## Optimization of LH- or RH-S$x$/Ag structures

SERS efficiency of the created S$x$/Ag structures was evaluated using an achiral model SERS probe – rhodamine 6G (R6G) (Fig. 1(c), (d), and Fig. S13, ESI†). Two basic parameters were considered: (i) the structure of SERS-responsive silver layer ("apparent" thickness and roughness, adjusted by silver deposition time) and (ii) the shape of the shurikens (Fig. S2 and S4, ESI† and related discussion). We used 785 nm as the Raman excitation wavelength and, as mentioned earlier, the illumination of pristine Au shurikens does not produce any intense SERS signals (Fig. 1(c) bottom graph). The Au only shuriken nanostructures most likely produces weak plasmonic energy confinement as the incident photon is "smeared" across the shuriken structure (Fig. S5, ESI†), with a size comparable to the excitation wavelength. In other words, the incident photons can be absorbed even in this case, but the evanescent field of the excited chiral plasmon is restricted by the relatively large space (determined by each shuriken unit geometry). This disagrees with the common plasmonic-based SERS requirement, where the photon energy should be focused in a plasmonic hot spot within 1–2 nm dimension. In turn, silver deposition led to the generation of plasmon active silver clusters which provides the plasmonic hot spots required for SERS (Fig. S14, ESI†). After the silver deposition, we observed an intense SERS response from R6G molecules, which have a similar trend on both kinds of shurikens rotations (Fig. 1(c), (d) and Fig. S13, ESI†).

Since the Ag cluster geometry and gaps between clusters are determined by the deposition time, there is an increase in R6G intensity with deposition time (up to $1.3 \pm 0.8$ nm of "effective" thickness) followed by the decrease as deposition time is prolonged as the clusters are likely merging with longer deposition times. This tendency was observed for all shuriken geometries (S1–S6), but was especially pronounced in LH-S3/Ag or RH-S3/Ag cases. The most enhanced response was in case of 1.3 nm Ag and was therefore chosen as the optimal "averaged" thickness. In addition to deposition time, shuriken geometry significantly affects the absolute SERS efficiency for all thicknesses (Fig. 1(c) and (d)). In particular, the SERS enhancement factor (EF) on S5/Ag and S6/Ag was less than $1.2 \times 10^3$; it was moderate on S4/Ag and S2/Ag shuriken types ($2.3 \times 10^3$ and $3.3 \times 10^3$); and the highest EF = $3.9 \times 10^3$ in the case of S3/Ag structures. The trend in the EF changes are aligned with changes in the optical response (Fig. S4, ESI† and related discussion). The S3/Ag samples showed more pronounced shifts in the reflection (absorption) spectra and increased absorption. Furthermore, the coupling of two plasmon active nanostructures commonly results in the changes of their plasmon absorption bands and optical response.[47] We observe a similar effect, more noticeable for S3/Ag samples, which also corresponds well with the presence of some plasmonic coupling between shuriken arrays and Ag clusters. Additionally, our simulations (Fig. S14 and S15, ESI†) also show strong electromagnetic field enhancements when the Ag nanoparticles are in the near vicinity of the high intensity fields generated by the shuriken but not when further away. The applied SERS excitation wavelength of 785 nm is also close to the S$x$/Ag resonance (Fig. S4, ESI†) and is far away from the common wavelengths (near 530 nm) used for SERS measurements on silver nanostructured layers. Deposition of R6G molecules on Ag (1.3 nm)/Si and SERS measurements with 785 nm excitation also did not produce any significant SERS responses (Fig. S12, ESI† and related discussion).

These observations and SERS EF results leads to the conclusion that the Ag nanoparticles and the Au shuriken nanostructures are coupled together to produce the large SERS enhancements. We can conclude that the Ag hot spots are primarily excited by the shuriken near field, but not by the incident photons. The silver dipoles are exposed to the super-chiral fields with enhanced chiral asymmetry, which can potentially result in higher sensitivity of chiral interactions and effects by local electric field and chiral dielectrics near the plasmonic boundaries. As a result, a significant difference in the Raman scattering spectra of small (cysteine) or large (DNA) biomolecular enantiomers could be expected in this case. Considering the SERS, reflectance and ORD measurements results, the (LH/RH)-S3/Ag (1.3 nm) structures were chosen for all further enantioselective SERS measurements.

## SERS-based discrimination of small molecules – L/D cysteine case

Following the above assumption, we proceeded to deposit L/D-cysteine on LH- or RH S3/Ag structures for SERS measurements. Considering that only molecules close to plasmonic hot spots can produce chiral SERS signal, we spin-coated cysteine solutions to generate thin films using varying cysteine concentrations to check the effect of probe analyte quantities on the plasmonic surface. Conditions of spin-coating were chosen to form a homogeneous, amorphous layer of cysteine (Fig. S16–S19, ESI† and related discussions). Structure S3/Ag produced the most intense SERS signals, however, we additionally verified that structure S3/Ag is the optimal one by comparing the signal intensity of L-cysteine ($10^{-4}$ M) measured on S1–S6 (Fig. S20, ESI†). The trend in the signal enhancement was similar to that presented in Fig. 1(d) for R6G, with overall lower intensity due to low cross section of cysteine molecules.

Results of L- or D-cysteine enantiomers measurements performed in common SERS mode (with utilization of non-polarized light and polarization-insensitive detector) are presented in Fig. 2. The SERS spectrum is identical to that

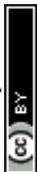

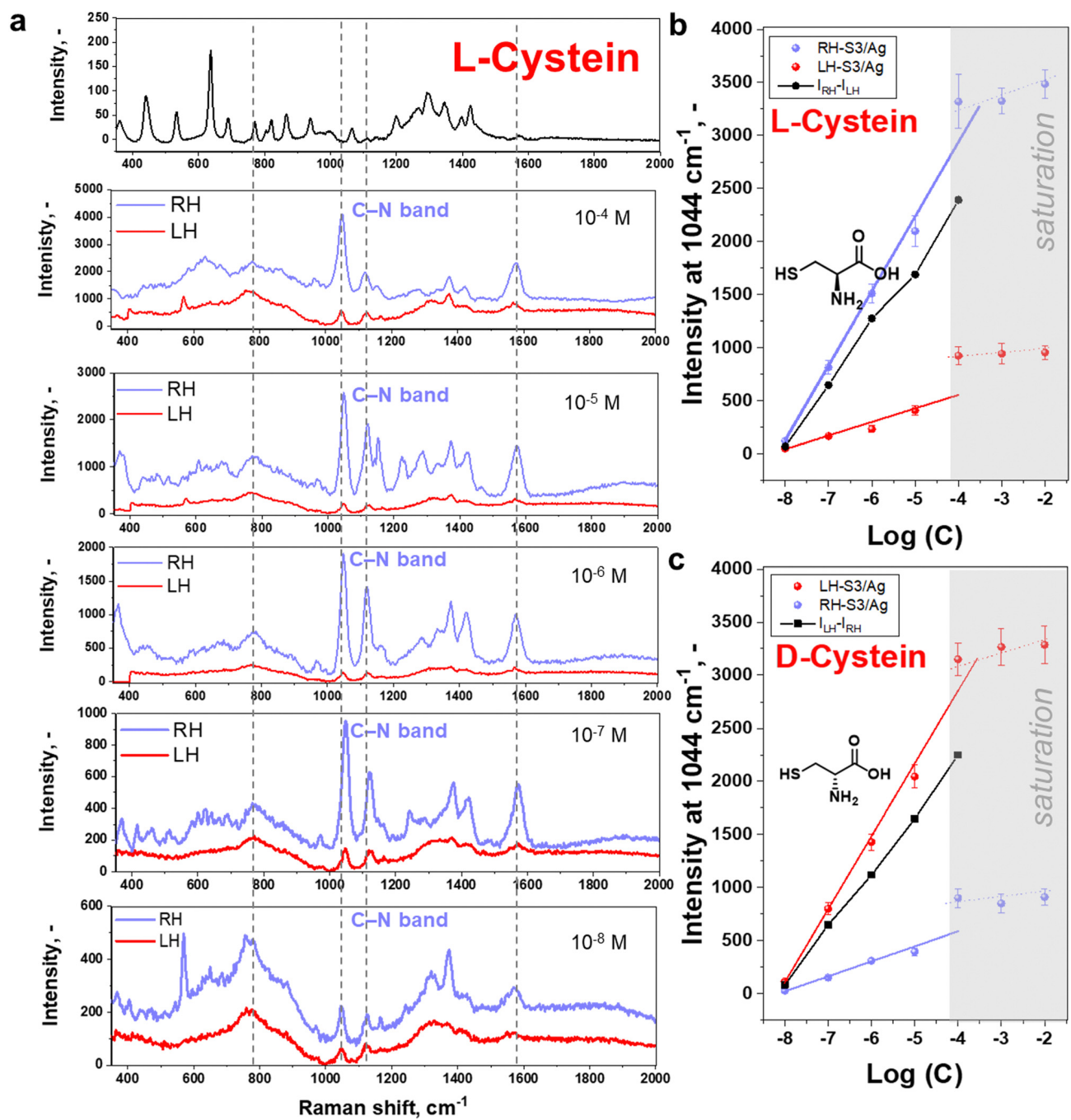

Fig. 2 (a) Stacked SERS spectra of L-cysteine enantiomer deposited on LH- or RH-S3/Ag substrates from methanol solutions with various initial concentrations ($10^{-4}$–$10^{-8}$ M range) measured at 785 nm; (b) and (c) the characteristic L-cysteine and D-cysteine SERS band (1044 cm$^{-1}$) intensity as a function of the initial enantiomers concentrations and plasmonic substrate LH- or RH-kind.

published previously for cysteine in the vicinity of Ag coated Au nanoparticles (comparative results are given in Table S1, ESI†).[48–51] The slight shift of peak position is likely due to the different excitation wavelengths and the interaction of plasmonic substrate with the cysteine *via* sulphur, amino and carboxylic acid groups.[50,51] In our case, however, the intensities of measured spectra are significantly affected by the enantiomers and shuriken handedness (Fig. 2(b) and (c)), RH- or LH-S3/Ag. The SERS bands were observed to be more pronounced in the case of L-cysteine deposited on RH-S3/Ag structures in comparison to D-cysteine enantiomers deposited on the same substrate. On the other hand, D-cysteine enantiomer SERS spectra are more intense in the case of LH-S3/Ag plasmonic substrate (Fig. S21, ESI†). The significant differences in characteristic SERS bands intensity (at 1044, 1579, and 1118 cm$^{-1}$) are observed from $10^{-8}$ M up to the $10^{-4}$ M initial concentration of cysteine enantiomers (Fig. 2(b), (c), and Fig. S22, ESI†). In the case of cysteine concentrations higher than $10^{-4}$ M (Fig. S23, ESI†), there was saturation of SERS signal intensity. Generally, the observed dependence of the SERS intensity from L/D-cysteine SERS intensity on the handedness of the plasmonic substrate (Fig. 2 and Fig. S21, S23, ESI†) clearly indicates that the coupling of Au shurikens and SERS active Ag layer is applicable for enantiomers discrimination up to $10^{-8}$ M using a standard Raman spectrometer.

To further verify the observed difference in enantioselective excitation by coupled chiral plasmonic structures, we investigated the homogeneity of intensities for characteristic cysteine-related peak at 1044 cm$^{-1}$ on LH- and RH-S3/Ag structures (Fig. 3). In this case we observed that the appropriate level of D-cysteine SERS response is higher by a factor of ~5 for LH-S3/Ag substrate compared to RH-S3/Ag (in latter case the inhomogeneous region on the right map rather refers to background deviations) (Fig. 3(a)). Even though cysteine is deposited over the entire surface of LH- and RH-S3/Ag samples, there is no SERS effect on the residual parts of the initially flat Au film (Fig. 3(b)). The dramatic difference in SERS mapping intensities further confirms the crucial role of the chiral plasmonic substrate. So, specifically designed Au shuriken – Ag nanocluster substrates can enable enantiomeric discrimination with a common Raman spectrometer with reliable sensitivity.

### Impact of SERS excitation wavelength

To better understand the coupled SERS structure and the importance of shuriken chiroptical properties (including optical rotation), we measure the impact of SERS excitation wavelengths on chiral SERS efficiency. The pristine S3 shuriken reflectivity measurements show an apparent plasmon resonance band near 750 nm, which is slightly enhanced and broadened after Ag layer deposition (Fig. S4, ESI†). When considering ORD measurements, Fig. S4b (ESI†), we observe that shuriken structures S5 and S6 show less intense ORD in comparison to the other structures. This difference in ORD can explain the poorer enantioselective SERS performance on LH- and RH-S5,6/Ag substrates. In addition, chiral response is

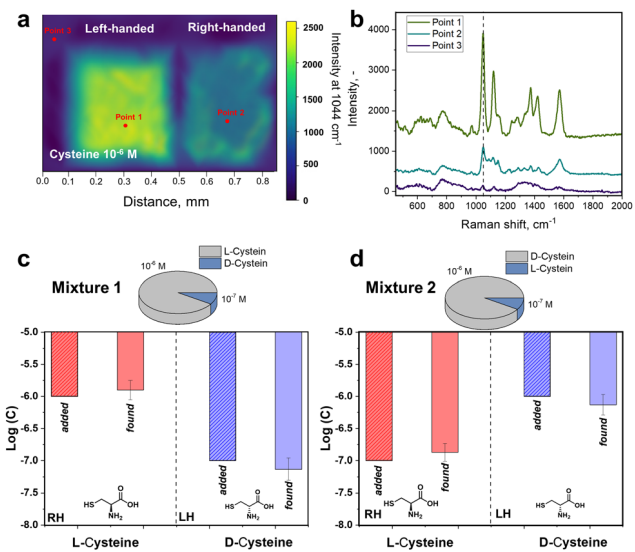

**Fig. 3** (a) SERS mapping of D-cysteine ($10^{-5}$ M) on RH- and LH-S3/Ag structures measured at 785 nm; (b) stacked SERS spectra from local points, designated on (a); (c) measured at 785 nm, (d) example of enantiomers content determination in L/D-cysteine mixtures using measurements on LH- and RH-S3/Ag SERS substrates.

also observed near 600 nm indicated by the ORD profiles, Fig. S4b (ESI†). Hence, utilization of standard Raman laser wavelengths near 600 and 750 nm could also result in the appearance of chiral response from the targeted analyte. However, Raman laser wavelength selection is typically limited by commercially available laser excitation sources and setups. Hence, Raman optical microscopes with 830 nm, 785 nm, 633 nm, and 532 nm sources that were available were selected for the SERS measurements. Note that the ORD signal for the first and last source is close to zero and we do not expect any similar response from it (Fig. S4 and S24, ESI†).

Impacts of SERS excitation wavelength on cysteine enantiomeric SERS response are presented in Fig. S25 (ESI†) and related discussion, which indicate that 785 nm is a suitable excitation wavelength for our structure. Using a 785 nm excitation, at which wavelength the RH-S3/Ag substrate has a positive value for the ORD (Fig. S24, ESI†), L-cysteine produces enhanced SERS signals. The same L-cysteine produces lower SERS signals on LH-S3/Ag substrates, which shows a negative value for the ORD at the same wavelength as the chiral responses are equal and opposite between the two opposite handed structures. In contrast, at 633 nm excitation, the SERS signal from D-cysteine is more pronounced from LH-S3/Ag where the optical rotation has a positive value unlike at 785 nm (Fig. S24, ESI†). Hence, this confirms that SERS spectra of enantiomer(s) measured at different wavelengths, will invert the selective chiral SERS response due to inversion in chiral response generated at those two different wavelengths. It should be noted, that based on the cotton effect, the opposite bi-signate ORDs also represent two opposite CD (albeit of a Lorentzian shape) responses in the same region and when moving between 633 nm and 785 nm, the CD would be expected to flip for every structure, indicating a change in the chiral responses. This dependence establishes the link between enantiomeric selectivity and excitation wavelength and provides an additional degree of control in enantiomer recognition reliability. It should also be noted that the apparent difference in L-cysteine SERS response was reached solely with wavelengths corresponding to the shuriken plasmon absorption (not a wavelength commonly used for rough Ag surface excitation wavelength – 530 nm). This difference confirms our previous assumption that Ag hot spot excitation by shuriken superchiral near field plays a crucial role in SERS-based enantiomeric discrimination.

### Enantiomers mixture analysis and substrates re-utilization

Commonly analytical probes contain both enantiomers, where the enantioselective sensing become more challenging. Taking advantages of coupled Au shuriken–Ag nanocluster substrates, the chiral detection performance of RH- and LH-S3/Ag substrates was verified by the SERS analysis of racemic mixtures containing $10^{-5}$ M L-cysteine and $10^{-6}$ M D-cysteine or vice versa (Fig. S26, ESI†). The measured averaged peak intensities were compared to previously "found" enantiomer concentrations using equations (Fig. 2(b) and (c)) Int$_{1044}$ = 700 ± 30 × log($c$) + 5740 ± 232 ($R^2$ = 0.993) for L-cysteine and Int$_{1044}$ = 690 ± 30 × log($c$) + 5600 ± 229 ($R^2$ = 0.992) for D-cysteine (Fig. 2(c) and 3). "Found" concentration values converged with the added amount of L/D-cysteine within the experimental error. As is evident, the Au shuriken–Ag nanocluster substrates determine enantiomer concentration within ~98% accuracy even in case of minor concentration of the target enantiomer. Hence, the proposed approach allows the identification of even negligible enantiomer presence without preliminary racemic mixture separation or chiral extraction.

Furthermore, we addressed the possibility of regeneration of the RH- or LH-S3/Ag substrates after chiral sensing of L/D-cysteine, which is an important requirement in various applications.[52] Firstly, we performed a few cycles of substrates utilization/regeneration within one day and over several days where substrates were also compared when stored in air or vacuum. Obtained results, presented in Fig. S27 and S28 (ESI†) show that the characteristic cysteine bands fully disappear after the substrate regeneration by gentle ultrasonication in methanol.[53] Several subsequent cycles of substrate utilization/regeneration reveal that the substrates can be re-used for up to 5 cycles without significant loss of functionality. In case of longer storage between regeneration/utilization, the more pronounced decrease of the intensity of cysteine-related Raman peaks was observed (probably due to Ag oxidation on air, evident as the appearance of dark spots on samples surface). Hence, freshly prepared RH/LH-S3/Ag can be re-used multiple times for chiral SERS sensing and with intermediate storage under vacuum.

### Versatility of the proposed approach – additional chiral analytes

To demonstrate the versatility of the proposed approach, we also tested the possibility of detecting alternative analytes – chiral amino acids L/D-methionine and (R)/(S)-thalidomide (medication used to treat a number of cancers and graft-versus-host disease[54]) using RH/LH-S3/Ag SERS substrates (Fig. 4).

In case of L/D-methionine, the SERS spectra demonstrate characteristic peaks at 1423 cm$^{-1}$ (sciss CH$_2$), 1324 cm$^{-1}$ (bend asym NH), 1250 cm$^{-1}$ (CH rock), 1120 cm$^{-1}$ (C–C str, rock NH$_3$$^+$, str C–N), 775 cm$^{-1}$ (CH$_2$ rock), 673 cm$^{-1}$ (CS str) cm$^{-1}$ (Fig. 4(a)). The intensity of L-methionine was significantly higher on RH-S3/Ag compared to LH-S3/Ag and vice versa, D-methionine measurements provide significantly better SERS response on LH-S3/Ag substrates. The same tendency was observed for (R)/(S)-thalidomide (Fig. 4(b)). SERS measurements also reveal appearance of characteristic thalidomide peaks at 1778 cm$^{-1}$ (C=O vib), 1536 cm$^{-1}$ (N–H def, C–C str vib), 1332 cm$^{-1}$ (C–H def vib), 1165 cm$^{-1}$ (C–Hin plane vib), 973 cm$^{-1}$ (C–J out of plane vib), 731, 653 cm$^{-1}$ (C–N str vib), 565 cm$^{-1}$ (C–N str vib, At ring vib). An obvious and significant

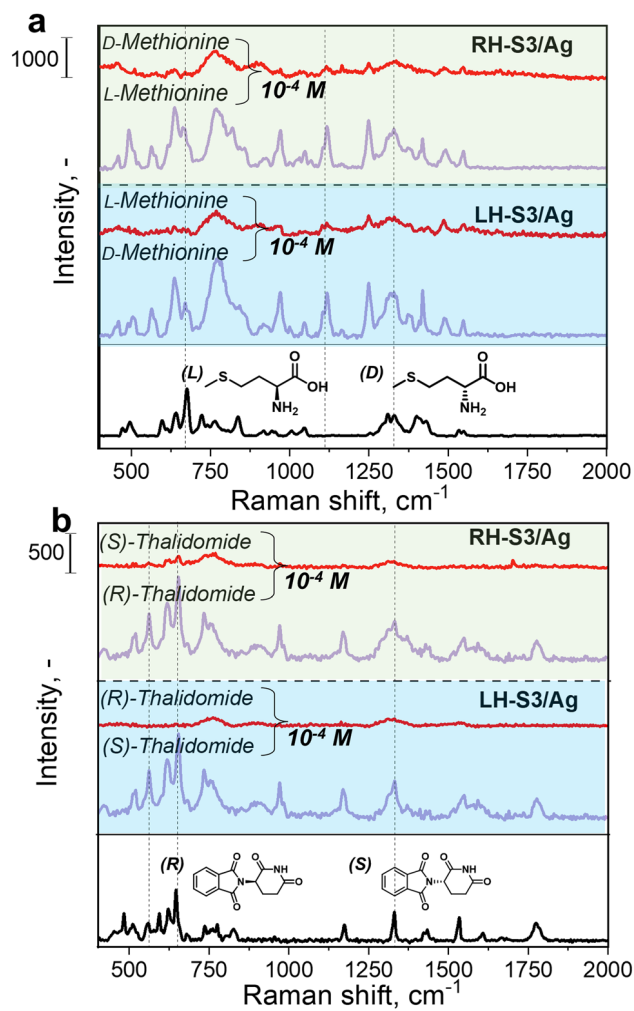

**Fig. 4** (a) Stacked SERS spectra of $10^{-4}$ M L/D-methionine enantiomer deposited on LH- or RH-S3/Ag substrates from methanol solutions measured at 785 nm, (b) SERS spectra of $10^{-4}$ M (R)/(S)-thalidomide enantiomer deposited on LH- or RH-S3/Ag substrates from methanol solutions measured at 780 nm measured at 785 nm. Bottom spectra (black) show Raman spectra of both compounds.

difference in SERS spectra of (*R*)- or (*S*)-thalidomide enantiomers on RH and LH-S3/Ag substrates was observed, confirming applicability for the analysis of chiral analytes with diverse structure. Hence, developed RH/LH-S3/Ag can be applied for the SERS detection of amino acids as well as for other chiral drugs.

### SERS-based discrimination of large biomolecules – DNA case

Utilizing chiral SERS for analysis of large biomolecules has fuelled our interest to apply RH/LH-S3/Ag for DNA analysis (DNA structure is in Fig. 5). We used double stranded DNA (this type of biomolecule commonly show CD response in the UV range)[55] consisting of 20 bases with the structure close to an ideal helical geometry with *ca.* 2 turns. We verified the possible increase of substrate RH/LH-S3/Ag temperature during SERS measurements, which can potentially cause innate denaturation of DNA. For this purpose, we grafted RH/LH-S3/Ag substrates with –C$_6$H$_4$–C≡N chemical moieties, which served as a molecular temperature probe (Fig. S29, ESI† and related discussion).[56] Comparison of measured SERS spectra collected at RT and elevated temperatures (external heating) indicated the insignificant increase of local substrate temperature under Raman laser illumination. Results of SERS sensing of DNA probe on LH- or RH-S3/Ag samples are presented in Fig. 5 (peak affiliation is given in Table S2, ESI†). The apparent spectral difference, significant SERS bands enhancement as well as their spectral separation was observed solely in the case of LH-S3/Au substrate, while the application of RH-S3/Au substrate did not produce any significant SERS signal.

To sum up, the proposed approach can be used for the analysis of both small organic molecules and large biomolecules. In constant to L/D-cysteine, where amino and carboxy-related Raman bands are enhanced, the surface enhancement of DNA SERS bands occurs homogeneously. This fact indirectly indicates that the molecular response can also be affected by the shuriken-induced polarization of the probe biomolecules, which further interact with plasmonic hot spots excited on silver clusters.

### Proposed mechanism of enantio-selective SERS enhancement

Our results demonstrate coupling of Ag-derived hot spots with chiral shuriken structures that produces an effective chiral SERS signal for enantiomer discrimination using a common Raman spectrometer. Based on the experimental results we propose several possible mechanisms. The first, one is the coupling of Ag dipoles and shurikens chiral structures, where silver hot spots are excited by the shuriken superchiral near field. In other words, the energy of incident photons (with 785 nm) is initially converted into a chiral plasmon (by shuriken(s)). The chiral plasmon of the shuriken pumps the energy into the Ag nanoparticles to generate a Ag plasmonic hot spot. In this case, the transition of superchirality from the shuriken to Ag hot spot can be expected. As a result, the probe molecules will be excited by a higher degree of local circular electric field polarization, which potentially ensures more resonant differences in SERS spectra of the probe enantiomers. Liu *et al.* proposed a similar concept using chiral Au nanofibers (Fig. S4, ESI†), unlike simpler dimers or nanorods, justifying the possibility of having a strong impact on the polarized excitation of the Ag dipoles.[32] We also perform an additional simulation of the behavior of Ag nanostructure (estimated as a single nanoparticle or two nanoparticles) subjected to shuriken plasmonic nearfields. Simulation results (Fig. S14, S15, ESI† and related discussion) indicate that under certain resonance conditions, the nanoparticle(s) is surrounded by relatively large chiral fields, which strongly modulate the behavior of the intrinsically achiral nanoparticles. Moreover, the excited near nanoparticle field is likely to be chiral in nature themselves and would affect the SERS response from the chiral molecules (taking into account the fact that even Ag nanostructures are the source of SERS). So, a difference in SERS efficiency (for various structures) and additional simulation results indicate that the plasmonic coupling may have a resonant nature which produces a localized superchiral field.

The second possibility is the overall enhancement of enantiomer spectra. This is determined solely by shuriken handedness and is independent on the probe molecule scattering tensor, concerning the position of optically active centers in the probe molecule structure. From this point of view, we can propose that the shuriken induces an overall polarization of analytes, which increases their Raman scattering coefficient in a similar way as it occurs in the case of metal-free SERS. Additional molecular polarization can provide the overall SERS intensity increase by a factor of $10–10^2$, which is in relatively good coincidence with the observed experimental data.[57] Still, the third explanation may consist in plasmon chirality transfer, occurring after the coupling of intrinsically non-chiral plasmon active nanoparticles (in our case – Ag clusters) and the chiral surrounding dielectric medium.[5,58] It could be expected that in our case (with the coupling of non-chiral plasmon active nanostructures and superchiral shuriken field) this phenomenon will be significantly enhanced and can ensure the increase of the enantioselective interaction of probe molecules with light, as was reported for common circular dichroism measurements. A fourth possible explanation is that the electric nearfield of the shuriken is affected by the chiral dielectric, effectively modulating the field intensity exciting the Ag clusters. The intensity and hence the SERS enhancement therefore

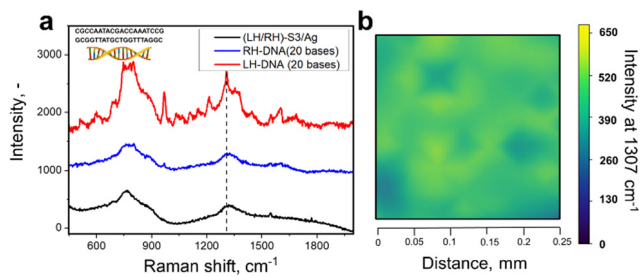

a

b

**Fig. 5** (a) Stacked SERS spectra of double-stranded DNA measured on LH- and RH-S3/Ag substrates measured at 785 nm, bottom spectrum (black) shows the ''background'' signal from pristine LH-S3/Ag substrate; (b) SERS mapping of double-stranded DNA obtained across the LH-S3/Ag substrate measured at 785 nm.

become dependent on the enantiomeric properties of the dielectric. However, to achieve a change in the electric field that is significant enough to cause the results demonstrated would require cysteine to have a chirality factor an order higher than considered practical.

At the moment, it is not possible to differentiate the exact mechanism for the observed enantioselective differences in SERS spectra. Moreover, the mechanisms described above can contribute simultaneously and synergistically. Further theoretical and experimental research in the field will be performed to elucidate the proper mechanism of the observed phenomena. However, it is clear that the proposed plasmon-active structures enable enantioselective SERS measurements using standard and inexpensive Raman spectrometers without the need of more complex ROA equipment. Moreover, the plasmon active substrates, including the shuriken support, are produced through scalable technology that can be considered as replaceable and disposable.[19] Hence, it will be possible to perform switching in Raman spectroscopy from normal to enantioselective measurement modes by exchanging the disposable plasmonic substrates.

## Conclusion

In this work, we demonstrated the key role of nanoarchitecture of plasmonic substrates in chiral SERS discrimination of small organic enantiomers and large biomolecules. The proposed design of chiral and SERS active substrates is based on the coupling of chiral plasmonic shurikens with array of Ag clusters. As a result, the Ag plasmonic cluster multimers are exposed to the superchiral fields close to shurikens, followed by expected excitation of circular polarization in plasmonic silver hot spots. This combination allows us to observe the high differences in the SERS intensities of chiral probed molecules (L/D-cysteine and DNA molecules) as a function of the chirality of the shuriken or the molecule. Optimization of this enantioselective detection/recognition was performed by varying (i) Ag layer thickness/roughness, (ii) geometry of shuriken, and (iii) selection of excitation wavelength, which should correspond to the chiral response of the coupled structure. These dependencies indicate the strong (and probably resonant) plasmonic coupling, which allows us to perform surface-enhanced, enantio-selective measurements with the wavelength-dependent enhancement of different enantiomers. Moreover, all measurements were performed in several minutes using a common Raman spectrometer and are also used to determine the enantiomeric composition in racemic mixtures. We demonstrated that a specific design of the plasmonic substrate with appropriately selected excitation wavelength could ensure the generation of nonlinearly polarized electric fields inside plasmonic hot spots triggering and enantiospecific SERS response from chiral molecules. This was experimentally confirmed by the dependency of the chiral selective SERS response on the excitation wavelength and the chiral response of the shurikens, supported by numerical simulations. The developed plasmon-active substrate can ensure the enantioselective measurements using a common Raman spectrometer without the necessity of a chiral light source or chiral sensitive detectors. Such a substrate and methodology are promising for rapid and simplified enantiomeric discrimination in pharmaceutical and chemical industrial applications.

## Experimental

Preparation and characterization of Sx/Ag structures are described in ESI.† Briefly, the template-assisted technique was used for the creation of Au Shurikens, followed by vacuum sputtering of silver. Surface morphology and metals distribution were checked using AFM and SEM-EDX measurements. Polarization-dependent reflectivity was determined using homemade equipment (see ESI† for detail). Raman spectra were recorded using a ProRaman-L and inVia spectrometers (530, 630, 780, and 850 nm excitation wavelengths). Spectra data collection was performed at the same experimental conditions (laser power, collection time), optimized for each used wavelength. From each sample, 10 random points were analysed on each shuriken array and all spectra were baseline corrected and smoothed using an 11-point averaging smoothing algorithm to reduce the baseline variability in the region between 450–2000 cm$^{-1}$, using Omnic Professional Software Suite (Thermo Scientific, Inc., Madison, WI). Deposition and preparation of probe analytes (cysteine enantiomers and DNA molecules) are also described in ESI.† Simulations were performed using COMSOL v5.6, a software for simulating physics using finite element method. Periodic boundary conditions were used to emulate the array of nanostructures. Perfectly matched layer conditions were used above and below the input and output ports. Linearly polarized EM wave was applied at normal incidence onto the structure.

## Author contributions

Olga Guselnikova – preparation and characterization of the materials, conduct of the experiments, design of the experiments, analysis of the data, and writing of the manuscript; Roman Elashnikov – conduct of the experiments, analysis of the data; Vaclav Svorcik – conduct of the experiments; Martin Kartau – preparation and characterization of the materials; Cameron Gilroy – preparation and characterization of the materials; Nikolaj Gadegaard – preparation and characterization of the materials; Malcolm Kadodwala – manuscript revision and proofreading; Affar S. Karimullah – study conception, characterization of the materials, simulations, and writing of the manuscript; Oleksiy Lyutakov – study conception, characterization of the materials, and writing of the manuscript.

## Conflicts of interest

There are no conflicts to declare.

## Acknowledgements

This work was financially supported by GACR under the project 20-19353S (OL) and ''Mega-grant'' Project No. 075-15-2021-585 of the Ministry of Science and Higher Education of the Russian Federation (AK). AK would like to acknowledge support by the UKRI & EPSRC (EP/S001514/1) and the James Watt Nano-fabrication Centre. AK, NG and MK acknowledge support from EPSRC (EP/S012745/1 and EP/S029168/1). We would also like to acknowledge Prof. Laurence Barron for his academic discussions and support.

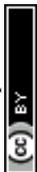